\documentclass[prx,twocolumn,superscriptaddress,nofootinbib]{revtex4-1}

\usepackage{amssymb}
\usepackage{amsmath}
\usepackage{bbold}

\usepackage{braket}
\usepackage{xcolor}
\usepackage{soul}
\usepackage{graphicx}
\graphicspath{{../figura1/}{../figura2/}}

\begin{document}


\title{Universal mean field upper bound for the generalisation gap of deep neural networks}

\author{S. Ariosto}
\affiliation{Dipartimento di Scienza e Alta Tecnologia and Center for Nonlinear and Complex Systems,
Università degli Studi dell'Insubria, Via Valleggio 11, 22100 Como, Italy}
\affiliation{I.N.F.N. Sezione di Milano, Via Celoria 16, 20133 Milano, Italy}

\author{R. Pacelli}
\affiliation{Dipartimento di Scienza Applicata e Tecnologia, Politecnico di Torino, 10129 Torino, Italy}

\author{F. Ginelli}
\affiliation{Dipartimento di Scienza e Alta Tecnologia and Center for Nonlinear and Complex Systems,
Università degli Studi dell'Insubria, Via Valleggio 11, 22100 Como, Italy}
\affiliation{I.N.F.N. Sezione di Milano, Via Celoria 16, 20133 Milano, Italy}

\author{M. Gherardi}
\affiliation{Università degli Studi di Milano, Via Celoria 16, 20133 Milano, Italy}
\affiliation{I.N.F.N. Sezione di Milano, Via Celoria 16, 20133 Milano, Italy}

\author{P. Rotondo}
\affiliation{Università degli Studi di Milano, Via Celoria 16, 20133 Milano, Italy}
\affiliation{I.N.F.N. Sezione di Milano, Via Celoria 16, 20133 Milano, Italy}

\begin{abstract}
Modern deep neural networks (DNNs) represent a formidable challenge for theorists: according to the commonly accepted probabilistic framework that describes their performance, these architectures should overfit due to the huge number of parameters to train, but in practice they do not.
Here we employ results from replica mean field theory to compute the generalisation gap of machine learning models with quenched features, in the teacher-student scenario and for regression problems with quadratic loss function. Notably, this framework includes the case of DNNs where the last layer is optimised given a specific realisation of the remaining weights. We show how these results -- combined with ideas from statistical learning theory -- provide a stringent asymptotic upper bound on the generalisation gap of fully trained DNN as a function of the size of the dataset $P$. In particular, in the limit of large $P$ and $N_{\textrm{out}} $ (where $N_\textrm{out}$ is the size of the last layer) and $N_\textrm{out} \ll P$, the generalisation gap approaches zero faster than $2 N_\textrm{out}/P$, for any choice of both architecture and teacher function. Notably, this result greatly improves existing bounds from statistical learning theory. We test our predictions on a broad range of architectures, from toy fully-connected neural networks with few hidden layers to state-of-the-art deep convolutional neural networks.
\end{abstract}

\maketitle

\section{Introduction}

In the last ten years deep neural networks (DNNs) \cite{Goodfellow-et-al-2016} revolutionised the field of Machine Learning, outperforming traditional methods in tasks that include image classification, speech recognition and time series prediction. Despite the enormous success in applications, the size of these architectures represents a puzzle for theorists. When Alexnet won the ImageNet competition in 2012 \cite{krizhevsky2012}, it had roughly 60 million parameters. In the following years, the VGG network delivered the state-of-the-art performance with more than 138 million parameters \cite{simonyan2014}. Nowadays, convolutional DNNs such as ResNet or Inception work by training about 10 million weights \cite{He_2016_CVPR, Szegedy_2015_CVPR}. 
According to common intuition, models with such a high number of degrees of freedom
should overfit the training data, and perform poorly on previously unseen data samples.
Statistical learning theory (SLT) \cite{Vapnik1999},
the established probabilistic framework to quantify the generalisation performance in
machine learning, does not provide any guarantee that such severely overparametrised
models should have any predictive power on test data \cite{Vapnik1999_2, Bousquet2004}.

Overcoming this conceptual puzzle engages
computer scientists, mathematicians and physicists alike \cite{Zdeborova2020,10.1145/3446776,doi:10.1146/annurev-conmatphys-031119-050745}.
In Ref. \cite{MeiE7665}, the authors provide a mean field view of the stochastic gradient dynamics of one-hidden layer networks by using the theory of gradient flows in Wasserstein spaces \cite{santambrogio2017}. Unfortunately, it is challenging to extend this approach to deeper networks. Other groups are studying the role of overparametrisation and related phenomena 
such as the double descent in the regime of lazy training \cite{pmlr-v119-d-ascoli20a, PhysRevX.10.041044, mei2019, pmlr-v119-20a, baldassi2021learning,PhysRevLett.127.278301,loureiro2021learning}. Also the statistical physics of kernel learning (originally started in \cite{PhysRevLett.82.2975}) has undergone a revival in the last few years \cite{canatar2021}, mainly due to the discovery of the Neural Tangent Kernel (NTK) limit of deep neural networks ---a mathematical equivalence between neural networks and a certain kernel that arises in the limit of large layer size \cite{10.5555/3327757.3327948, NEURIPS2019_0d1a9651}. Despite all these major conceptual advances in the field, it is fair to say that a unified framework to investigate and understand the generalisation performance of DNNs is still missing.

On the mathematical side, it is instructive to rationalise why the theorems proven in the framework of
statistical learning theory often yield very loose bounds when applied to practical problems (as brilliantly put forward in Ref.~\cite{Bottou2015} by Bottou or in the recent review \cite{belkin_2021} by Belkin). The goal of theorems in SLT is to provide \emph{distribution-independent uniform} bounds on the deviation between the generalisation and training errors. The formulation and the derivation of these theorems reveal a source of possible reasons for their poor quantitative performance: 
(i) empirically relevant data distributions may lead to smaller typical deviations than the worst possible case \cite{PhysRevResearch.2.023169,PhysRevLett.125.120601,PhysRevE.102.032119,e23030305, pastore_2021}; (ii) uniform bounds hold for all possible functions in the model, but better bounds may hold when one restricts the analysis to functions that perform well on specific (and significative) training sets.    

In this manuscript, we  build upon these considerations to develop a mean field theory for the generalisation gap (GG) of deep neural networks. Firstly, we employ non-rigorous but standard statistical physics tools of disordered systems \cite{mezard1987} to compute the generalisation and training errors of machine learning models with \emph{quenched features}, obtaining simple formulas in the regime of large training dataset size $P$.
In particular these results hold in the teacher-student scenario for a broad class of input-output distributions when the employed loss function is the mean squared error. 
Notably, this setup includes the case of DNNs where the $N_{\textrm{out}}$ weights in the last layer are optimised given any specific instance of the remaining weights.
Analogous results have been derived in the recent literature \cite{10.2307/26542784, canatar2021, loureiro2021learning}; here we show how to employ them to derive a 
universal mean field upper bound for the generalisation gap of fully trained DNNs. In the limit $N_{\textrm{out}} \ll P$ 
(a condition satisfied by most state-of-the-art DNNs,
even in the overparametrised regime where $P$ is small compared to the
total number of weights), 
a simple asymptotic upper bound emerges,
according to which the gap should approach zero faster than $2N_{\textrm{out}}/P$. This is our central result; in the large $P$ limit, greatly improves existing estimates. Finally, we check the validity of our mean field bound against several 
synthetic and empirical
data distributions and across a variety of different architectures, ranging from toy
DNNs with few fully-connected layers to state-of-the-art ones employed for challenging computer vision problems.

Although these results lacks the mathematical rigor of formal theorems, it takes a concrete step towards understanding why overparametrised DNNs work in practice, and may guide to the formulation of more informed and accurate bounds for the generalisation gap of modern DNNs.


\section{Generalisation gap of quenched features models}

We start by briefly describing the setting of the supervised learning problem that we will study throughout the manuscript. Let us consider a training set $\mathcal T$ made of $P$ independent identically distributed (IID) random observations,
$\mathcal T = \left\{\left(\mathbf{x}^{\mu},y^\mu\right)\right\}_{\mu=1}^P$,
where the $\mathbf x^\mu$'s are $D$-dimensional vectors drawn by an input probability distribution $\rho (\mathbf x)$ and the $y^\mu$'s are scalar outputs provided by a real-valued teacher function $f_{\textrm{T}}$, i.e.~$y^\mu = f_{\textrm{T}} (\mathbf x^\mu)$. Under these assumptions, the joint input/output probability distribution $\rho_{\textrm{I/O}}(\mathbf x, y)$ is given by $\rho_{\textrm{I/O}} (\mathbf x, y) = \rho (\mathbf x) \delta \left(y - f_{\textrm{T}} (\mathbf x)\right)$.

Our first goal is to compute the generalisation performance of a model (the so-called {\it student}) of the following form 
\begin{equation}
f_{\textrm{S}}(\mathbf x) = \sum_{\alpha = 1}^N v_\alpha \phi_\alpha (\mathbf x) = \mathbf{v} \cdot \pmb{\phi} (\mathbf x),
\label{student}
\end{equation}
where $\pmb \phi$ is an $N$-dimensional feature map and $\mathbf v$ is an $N$-dimensional vector of real weights to be optimised.
The average generalisation and training errors are defined as:
\begin{align}
\label{epsilong}
\epsilon_{\textrm g} &= \left\langle\int d^D x \rho(\mathbf x) \left[f_{\textrm{T}} (\mathbf x) - f^*_{\textrm{S}} (\mathbf x)\right]^2\right\rangle_{\mathcal T}\,, \\
\label{epsilont}
  \epsilon_{\textrm t} &=  \left\langle \frac{1}{P} \sum_{\mu=1}^P \left[f_{\textrm{T}} (\mathbf x^\mu) - f^*_{\textrm{S}} (\mathbf x^\mu) \right]^2 \right\rangle_{\mathcal T}\,,
\end{align}
where $\braket{\cdot}_{\mathcal T}$ indicates the average over all the possible realisations of a training set of size $P$, and the optimised function $f^*_{\textrm{S}}$ corresponds to the choice of the vector $\mathbf v^*$ that minimises the quadratic loss
  $P^{-1} \sum_{\mu=1}^P \left[f_{\textrm{T}} (\mathbf x^\mu) - f_{\textrm{S}} (\mathbf x^\mu) \right]^2$ for a given instance of the dataset $\mathcal T$.
Although in principle this approach can be developed for arbitrary loss functions \cite{loureiro2021learning}, here we will only consider quadratic loss
and regression problems, which are considerably simpler to deal with analytically \cite{canatar2021, pmlr-v139-mel21a,Coolen_2020}.

The generalisation power of a machine learning model can be measured by its {\it generalisation gap},
\begin{equation}
\label{epsgap}
\Delta \epsilon = \epsilon_{\textrm g} - \epsilon_{\textrm t}\,,
\end{equation}
which expresses the average performance difference of a trained model 
between its training dataset and unseen data drawn from the same distribution.
Crucially, it is possible to express the generalization and training errors as a function of the features; as we will discuss in the following, this ingredient is fundamental to provide insight on the generalisation gap of fully-trained DNNs.
Here the calculation of the average generalisation and training errors is performed using the well-known replica method, a standard statistical physics technique developed to study disordered systems \cite{mezard1987}.
Optimisation is addressed introducing an effective Hamiltonian given by the sum of the training loss with a regularisation term: 
\begin{equation}
\mathcal L = \frac{1}{2}\sum_{\mu=1}^P \left[f_{\textrm{T}} (\mathbf x^\mu) - f_{\textrm{S}} (\mathbf x^\mu) \right]^2 + \frac{\lambda}{2} \sum_{\alpha=1}^N (v_{\alpha})^2\,,
\label{MSEloss}
\end{equation}
gauged by the regularisation parameter $\lambda>0$.
Given the convexity of our problem, we can make use of the replica symmetric ansatz, which is known to deliver the correct result for convex optimisation when
$P \gg 1$ \cite{PhysRevLett.82.2975}. 
The ground state of the effective Hamiltonian, given by the optimised $f^*_{\textrm{S}}$, is finally evaluated from the replicated partition function in the large $P$ limit via the saddle point method. 
This approach is rather similar to that developed for the random features model (RFM) \cite{PhysRevX.10.041044} and for kernel regression \cite{canatar2021} and it is discussed in detail in the appendices.

It turns out that the analytical expressions for the generalisation and training errors depend on the following integrals over the input probability distribution:
\begin{equation}
    \begin{split}
        J_\alpha &= \int d^D x \rho(\mathbf{x}) f_{\textrm{T}}(\mathbf{x}) \phi_\alpha (\mathbf{x})\,, \\ 
         \Phi_{\alpha \beta} &= \int d^D x \rho (\mathbf{x}) \phi_{\alpha} (\mathbf{x}) \phi_\beta (\mathbf{x})\,, \\
        T &= \int d^D x \rho (\mathbf{x}) f^2_{\textrm{T}}(\mathbf{x})\,.
    \end{split}
    \label{JphiT}
  \end{equation}
  with $\alpha,\beta=1,\ldots N$.
The vector ${\bf J}$ and the matrix $\mathbf{\Phi}$ depend on the specific choice of the feature map $\pmb{\phi}$ and respectively represent a teacher-feature and a feature-feature overlap, whereas the scalar quantity $T$ is by definition the trivial predictor \cite{10.1214/20-AOS1990}
of the regression problem and it provides a natural scale to compare different learning problems. Using these definitions, we find the following compact representation for the generalisation and training errors (valid in the thermodynamic limit of large $D,N,P$ as discussed in appendix \ref{AppA}):

\begin{equation}
    \begin{split}
    \epsilon_{\textrm{g}} &= \frac{\epsilon_{\textrm{g}}^{\textrm{R}} + (\kappa \lambda)^2 {\bf J}^T \mathbf{\Phi}^{-1} {\bf G}^{-2}{\bf J}}{1-P \textrm{Tr}\left(\mathbf{\Phi}^2 { \bf G}^{-2}\right)}\,, \\ 
    \epsilon_{\textrm{t}} &= \frac{\epsilon_{\textrm{g}}^{\textrm{R}}}{\kappa}+ \frac{\epsilon_{\textrm{g}}}{\kappa}\left(\frac{\kappa-1}{\kappa}-\frac{N}{P}\right)\,, \\ 
    \end{split}
\label{traintesteq}    
\end{equation}
where the matrix ${\bf G} = \kappa\lambda \mathbb{1} + P \mathbf{\Phi}$ is invertible and the variable $\kappa$ is self-consistently defined via the following equation:
\begin{equation}
    \kappa = 1+ \kappa \textrm{Tr} \left(\mathbf{\Phi} {\bf G}^{-1} \right)\,.
\end{equation}
The residual generalisation error $\epsilon_{\textrm{g}}^{\textrm{R}}$ corresponds to the best possible performance of the quenched model on the dataset, under the assumption of full knowledge of the input/output probability distribution, and it is given by:
\begin{equation}
\label{epsR}
    \epsilon_{\textrm{g}}^{\textrm{R}} = T -{\bf J}^T \mathbf{\Phi}^{-1}{\bf J}\,.
\end{equation}

It is worth notincing that for strictly infinite size of the dataset $P$, $\kappa \rightarrow 1$ and it is easy to prove that $\epsilon_{\textrm{g}} \rightarrow \epsilon_{\textrm{g}}^{\textrm{R}}$  and $\epsilon_{\textrm{t}} \rightarrow \epsilon_{\textrm{g}}$, which provide a first consistency check of the validity of the mean field theory. Additionally, the self-consistent definition of $\kappa$ and the way it enters in the expression for the generalisation error are the same as in the recent work on kernel regression by Pehlevan's group \cite{canatar2021}. This should not come as a surprise, since one could specialise the general quenched features that we employ here to the case of polynomial, Gaussian or NTK kernels 
. For these particular choices of the quenched features, Eq. (\ref{traintesteq}) just provide a different representation of the generalisation error formula given in \cite{canatar2021}. A generalization of Eq. (\ref{traintesteq}) has been recently proved in \cite{loureiro2021learning}.

Starting from Eq. \eqref{traintesteq} it is possible to perform an asymptotic analysis for large size of the training set $P$. This is particularly simple if we assume that both $P$ and $N$ are large, $N \ll P$ and the regularisation parameter $\lambda$ is finite. In this case we easily obtain that $\kappa \sim 1+ N/P$ and $\mathbf{G}\sim P \mathbf{\Phi}$. Using these asymptotic expressions,
 and normalising by the natural scale of the problem -- i.e.~the trivial predictor $T$ defined in Eq. \eqref{JphiT} -- 
we can compute the normalised generalisation gap:
\begin{equation}
\Delta \tilde\epsilon\equiv \frac{\Delta \epsilon}{T}\simeq 2 \frac{\epsilon_{\textrm{g}}^{\textrm{R}}}{T} \frac{N}{P} \,.
\label{asymptoticgap}
\end{equation}
Note that Eq. \eqref{asymptoticgap} does not necessarly imply a linear scaling with $N$ since $\epsilon_{\textrm{g}}^{\textrm{R}}$ may still retain a dependence on  $N$ (as implied from Eq. \eqref{epsR}).

\section{Generalisation gap of fully trained DNNs}

Let us now suppose that the quenched features of the model under consideration are given by a DNN. For instance, in the special case of a fully-connected architecture with one hidden layer, we have that the function implemented is $f_{\textrm{1HL}} (\mathbf x)= \sum_{\alpha=1}^N v_{\alpha} \sigma \left(W_{\alpha} \cdot \mathbf x\right)  $, where for simplicity we have set all the biases to zero. The $W_\alpha$'s are the $D$-dimensional vector weights of the hidden layer and $\sigma$ is a generic well-behaved activation function. Here the quenched features are given by $\phi^{\textrm{1HL}}_\alpha (\mathbf x, W) =  \sigma \left(W_{\alpha} \cdot \mathbf x\right)$. More in general, let us consider a DNN with a fully-connected last layer. The specific architecture of the first layers is uninfluential. This class includes all the relevant state-of-the-art architectures. Let us fix the dimension of the last layer to
$N = N_{\textrm{out}}$ and let us split the weights $\vartheta$ of the network as $\vartheta = \{ \mathbf v, \mathcal W\}$ where $\mathbf v$ is the vector of $N_{\textrm{out}}$-dimensional weights of the last fully-connected layer, whereas $\mathcal W$ is a short notation for all the remaining weights of the DNN. We introduce the feature map notation $\phi^{\textrm{DNN}}_{\alpha} (\mathbf x, \mathcal W)$ to indicate the corresponding quenched features.

\begin{figure*}[th]
\includegraphics[width=1\textwidth]{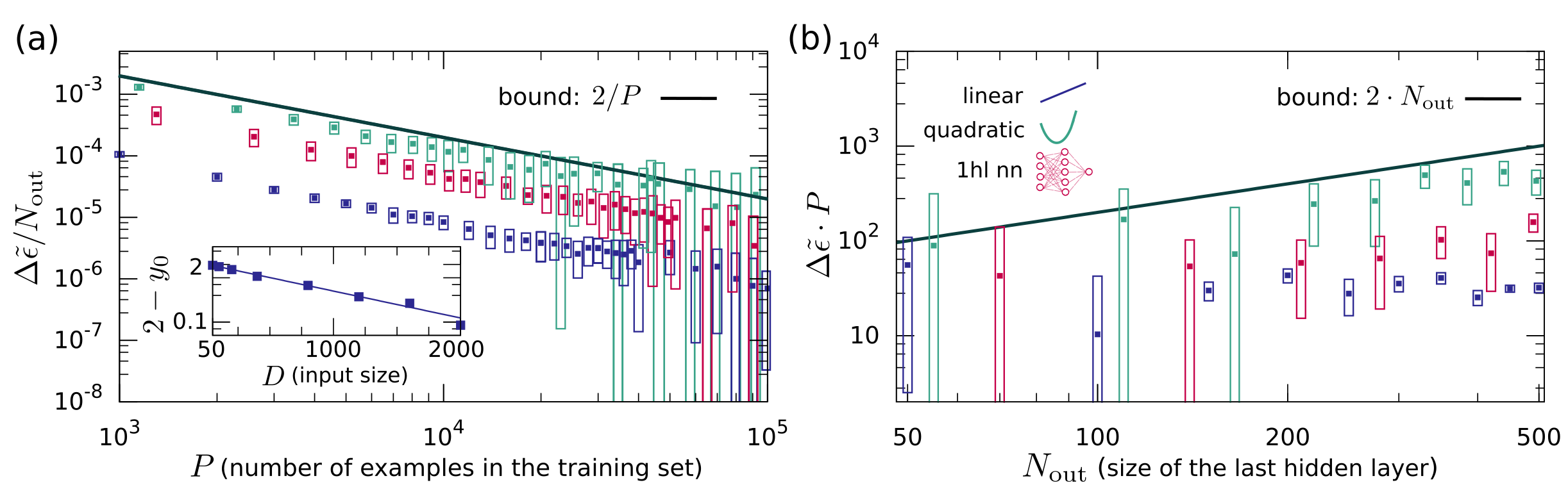}
\caption{\textbf{Generalisation gap in the lazy-training regime.} The behaviour of the normalized GG for a one-hidden layer student architecture is displayed for the three different classes of teacher outlined in the main text: linear (blue symbols), quadratic (green) and one-hidden layer (red). The solid black line marks our mean field upper bound.
(a) Normalised GG (rescaled by $N_{\textrm{out}}$) as a function of the training dataset $P$.
Data points are the result of an average over $50$ different realisations of the teacher and of the input (of dimension $D = 50$) with $N_{\textrm{out}} = 400$. We observe that the functional form of the rescaled GG $\Delta \tilde{\epsilon}$ is compatible with $y_0/P$ over two decades. In the inset we consider different input dimension for the linear teacher case, showing that by increasing $D$, the residual generalisation gap converges exponentially fast to the bound, i.e.
the prefactor $y_0$ (which is obtained by a standard fitting procedure) converges exponentially from below to $2$ as $D\rightarrow \infty$ with a rate $\gamma \approx 0.0014$ (see also appendix \ref{AppB}). (b) The normalised GG (multiplied by $P$) is shown as a function of the size of the hidden layer $N_{\textrm{out}}$. Simulations are performed with $P$ equal to $2\cdot 10^4$ (linear teacher), $4\cdot 10^4$ (quadratic) and $8\cdot 10^4$ (1HL), and by averaging over $20$ different teacher and input realisations ($D=50$). Error bars in both panels correspond to one standard error. Typical (rescaled by $T$) training errors in the lazy-training regime are of order $10^{-1}$ and are systematically smaller for the linear teacher.}
\label{fig:1}
\end{figure*}

We now reconsider the results provided by Eq. \eqref{traintesteq} when specialised to the quenched features of a DNN. The crucial observation is that mean field theory provides the average generalisation and training errors for \emph{each} realisation of the weights $\mathcal W$. In other words, given a specific configuration $\bar{\mathcal W}$, our theory predicts the corresponding average generalisation and training errors, supposing that the weights $\mathbf v$ of the last layer are set to the optimal value that minimises the training loss at fixed $\bar{\mathcal W}$. From now on, we use the notation $ \epsilon_{\textrm{g}} \left(\mathcal W\right)$, $\epsilon_{\textrm{g}}^{\textrm{R}} \left(\mathcal W\right)$, $\epsilon_{\textrm{t}} \left(\mathcal W\right)$ to stress that the generalisation and training errors depend on $\mathcal W$ via the teacher-feature ${\bf J}$ and feature-feature $\mathbf{\Phi}$ overlaps.

Therefore, the result in Eq. \eqref{asymptoticgap} holds for each  given realisation of the weights $\mathcal W$ of the DNN if we assume perfect training over the last layer.
See also the recent conjecture put forward in \cite{loureiro2021learning}.
In particular, this equivalence holds for a fully trained configuration 
$\theta^*\equiv \{v^*, \mathcal W^*\}$ that is a {\it local} minimum of the loss. Unfortunatley, such local minimum may depend on the size $P$ of the training set, and so it does $\epsilon_{\textrm{g}}^{\textrm{R}} \left(\mathcal W\right)$.
However, since the residual generalisation error (\ref{epsR}) is positive by definition and bounded by $T$, it follows that $0 \leq \epsilon_{\textrm{g}}^{\textrm{R}} \left(\mathcal W\right) /T \leq 1$ for every $\mathcal W$. As such, this provides us with the following asymptotic mean field upper bound for the (normalised) generalisation performance of a DNN:
\begin{equation}
\Delta \tilde{\epsilon} \left(\mathcal W\right) \leq  \frac{2 N_{\textrm{out}}}{P}\,,
\label{meanfieldbound}
\end{equation}
which is the central result of this manuscript.

\section{Numerical experiments}

\subsection{Toy DNNs with synthetic datasets.} We start by testing our bound on a fully-connected architecture with one-hidden layer and ReLU activations. We have chosen three different teacher classes of increasing complexity: (i) a linear function $f_{\textrm{T}}(\mathbf x) = \mathbf t \cdot \mathbf{x}$; (ii) a quadratic polynomial $f_{\textrm{T}}(\mathbf x) = \mathbf t \cdot \mathbf{x} + (\mathbf t \cdot \mathbf{x})^2$; (iii) a fully-connected one-hidden layer (1HL) architecture with ReLU activations, $f_{\textrm{T}}(\mathbf x) = \sum_{\alpha =1}^M q_\alpha \textrm{ReLU}\left(\mathbf{S}_\alpha \cdot \mathbf x\right)$. 
See also appendix \ref{AppB} for details on these architectures and on the inputs choice.


\begin{figure*}[t]
\includegraphics[width=1\textwidth]{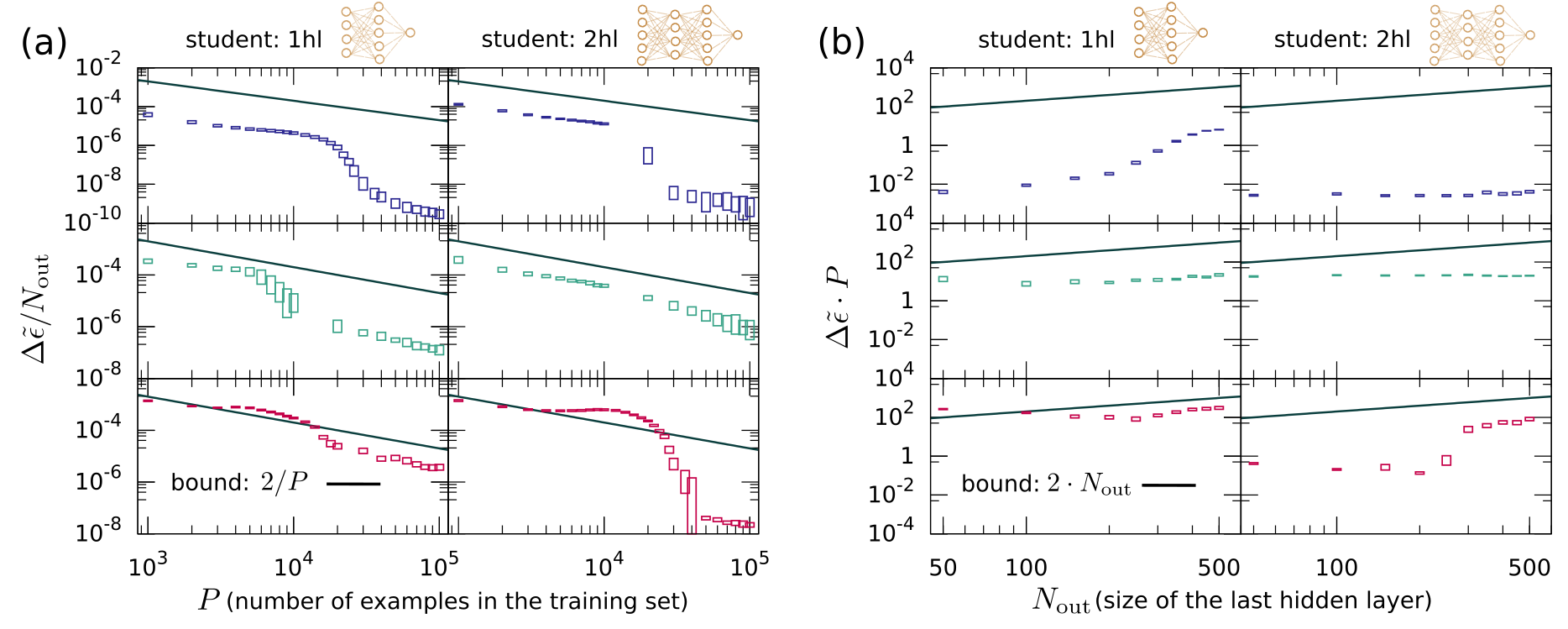}
\caption{\textbf{Generalisation gap of fully-trained toy DNNs} with one (left columns in (a)-(b)) and two (right columns) fully-connected hidden layers for the three different teacher classes outlined in the main text. From top to bottom: linear (blue symbols), quadratic (green) and 1HL teacher architectures (red).  The solid black line marks the mean field upper bound. (a) Normalised GG (rescaled by $N_{\textrm{out}} $ as a function of the size of the training set $P$.
Data points are the result of an average over $50$ realisations of the teacher and of the input ($D=50$) with $N_{\textrm{out}} = 100$. (b) Normalised GG (multiplied by $P$ vs. the size of the last hidden layer $N_{\textrm{out}}$. Data is averaged over $20$ different teacher and input ($D=50$) realisations with $P=4\cdot 10^4$. Error bars in both panels correspond to one standard error. Typical training errors are of the order of $10^{-6}$ for each teacher/student pair, except in the case 1HL/1HL where the training error is of order $10^{-3}$.}
\label{fig:2}
\end{figure*}

We first consider DNNs where only the last layer is trained and the remaining weights are kept fixed to their initialisation values $\bar{\mathcal W}$, i.e. we consider the lazy training regime. Since the $\bar{\mathcal W}$'s do not change during training, the residual generalisation error $\epsilon_{\textrm{g}}^{\textrm{R}} \left(\bar{\mathcal W}\right)$ is independent of $P$ and one
expects not only the bound \eqref{meanfieldbound} to hold but also the generalisation gap to scale precisely as $1/P$ for $P$ large enough. This is verified in Fig. \ref{fig:1}a for the three different teacher classes introduced above. On the other hand, as already noted, $\epsilon_{\textrm{g}}^{\textrm{R}} \left(\bar{\mathcal W}\right)$ may still retain a dependence on the last layer size $N_{\textrm{out}}$. In particular, one may expect that increasing $N_{\textrm{out}}$ will decrease the residual generalisation error, as this increases the number of functions available to approximate the target $f_{\textrm{T}} (\mathbf x)$. Therefore, for large $P$ and $N_{\textrm{out}}$ we expect $\Delta \tilde{\epsilon}$ to increase {\it at most} linearly as a function of $N_{\textrm{out}}$. The numerical behaviour of the GG as a function of $N_{\textrm{out}}$ is shown in Fig. \ref{fig:1}b. Once again, our mean field bound holds, but different scaling with $N_{\textrm{out}}$ can be appreciated. In particular, the GG is almost constant for the linear teacher, whereas it has an approximately linear behaviour for the quadratic one, reflecting different dependencies of the residual generalisation error from the last layer size. Note also that in both panels of Fig. \ref{fig:1} the GG is systematically lower for the linear teacher case, confirming the intuitive expectation that the linear problem should be the easiest to learn.

It is worth remarking that the input dimension $D$ only enters the theory through the residual generalisation error \eqref{epsR}. Interestingly, as one increases
the input dimension $D$, the normalised generalisation gap seems to saturate the unit bound with an exponential convergence
in $D$, as shown in the inset of Fig. \ref{fig:1}a for the linear teacher case (more details in appendix \ref{AppB}). Currently we have no theory for this.

We next consider fully-trained DNNs. Here the weights $\mathcal W$ are trained
and the residual generalisation error $\epsilon_{\textrm{g}}^{\textrm{R}} \left(\mathcal W\right)$ may dipend on the training set size $P$. 
Suppose for instance that there exists a configuration of the weights $\mathcal W^\dagger$ such that the residual generalisation error $\epsilon_{\textrm{g}}^{\textrm{R}} \left(\mathcal W^\dagger\right) = 0$, i.e.~the DNN can learn the teacher function perfectly. Intuitively, as the size of the dataset $P$ grows, we expect that the DNN will be capable of finding configurations $\mathcal W$ that are closer to the optimal one $\mathcal W^\dagger$. This means that the residual generalisation error will decrease in some way towards zero as a function of $P$ for $P$ large enough, and that the the GG will drecrease faster than $1/P$.

In the complementary case where the DNN is not capable of learning the target function,
there will be a non-zero residual generalisation error $\epsilon_{\textrm{g}}^{\textrm{R}} \left(\mathcal W\right) = \hat{\epsilon}^{\textrm{R}}$ even for $P\rightarrow \infty$: the $1/P$ scaling of the GG will thus be restored asymptotically.

\begin{figure}[th]
\includegraphics[width=0.45\textwidth]{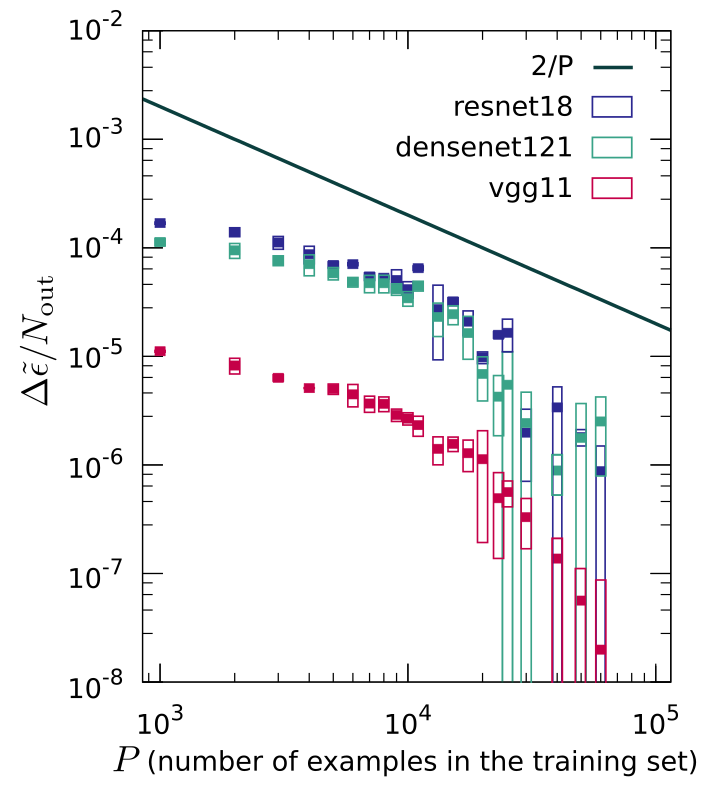}
\caption{\textbf{Generalisation gap for three state-of-the-art architectures trained on the MNIST dataset of handwritten digits.} The dependence of the normalised generalisation gap on the size of the training set $P$ is qualitatively similar for ResNet18 (blue symbols), DenseNet121 (green) and VGG-11 (red).  The solid black line marks the mean field upper bound.
Notice that by rescaling the GG by $N_{\textrm{out}}$ we can better compare architectures with different last layer size. Averages have been performed over three different initial conditions for the architecture weights, and error bars measure one standard error. Typical training errors are of order $10^{-4}$.
}
\label{fig:3}
\end{figure}

Numerical simulations for fully-trained DNNs are shown in Figure \ref{fig:2}. 
As expected from the considerations raised above, and
differently from the lazy training regime, here we do not observe a simple $1/P$ scaling of the gap (Fig. \ref{fig:2}a). On the contrary, the GG curves display two learning stages as a function of $P$,  with the GG falling systematically below the mean field bound above $P \sim 10^4$. Once we enter in the second learning stage (for $P$ of the order of $10^4$) the gap however seems to approach zero as fast as $1/P$ for both one
and two hidden layer architectures and across the different synthetic datasets, suggesting a fine constant residual generalisation error in this second learning stage.
In Fig. \ref{fig:2}b we analyse the generalisation performance as the width of the last layer $N_{\textrm{out}}$ grows. According to our predictions, the bound holds and a linear or sub-linear degradation of the generalisation performance is systematically observed across the different student and teachers architectures.

\subsection{State-of-the-art architectures.} As an additional and more challenging test, we present the results for the generalisation gap obtained by training
three different state-of-the-art convolutional architectures (ResNet18, DenseNet121 and VGG-11) on the MNIST dataset of handwritten digits \cite{MNIST}. Notice that this problem is in principle a classification problem, but our theory has been formulated for regression. For this reason we implemented the learning problem as a regression task: for each digit vector $\mathbf x$, the associated output is simply the integer number, between $0$ and $9$, corresponding to its class. 
Coherently, the performance of the network is not measured using the standard accuracy (i.e., the fraction of correctly classified digits), 
but as the mean square deviation
between the network's output and the class index.

A summary of the simulations is found in Fig. \ref{fig:3} (details on the learning protocols are provided in appendix \ref{AppB}). Remarkably, our bound is also satisfied in the case of state-of-the-art architectures trained on a dataset of practical relevance for computer vision. Moreover, notice that in the regime we are exploring the generalisation gap approaches zero faster than $1/P$.

\section{Discussion and Future Perspectives}

Our mean field analysis, while lacking the full rigor of theorems, establishes a much more stringent bound for the generalisation gap w.r.t. the ones obtained in the strict context of statistical learning theory. For instance, in the case of classification problems, SLT predicts an upper bound roughly proportional to $\sqrt{N_{\textrm{tot}}/P}$ with $N_{\textrm{tot}}$ being the {\it total} number of network parameters \cite{bartlett2019nearly} (more correctly $N_{\textrm{tot}}$ should be replace by the so-called Vapnik-Chervonenkis dimension of the DNN),
while our upper bound depends on the number of parameters of the network
only through the width of the last layer;
this may lead to speculate that the width of the last layer plays a special role
in DNN architectures.
The reason why we have been able to obtain this improved non-rigorous upper bound on the generalisation gap is
related to taking optimisation into account, at least in the last layer.
Whereas SLT attempts to find results for the generalisation gap that hold for 
every function in the model class, here we are restricting to special (but significative)
elements of the class.
In particular, for the specific case of a DNN with parameters $\vartheta = \{ \mathbf v, \mathcal W\}$, 
classic SLT bounds hold for every realisation of $\vartheta$. On the contrary, our approach assumes that at fixed $\mathcal W$ the weights $\mathbf v$ of the last layer are optimised w.r.t.~the training set. 

It is fair to stress the major limitations of our mean field bound: (i) crucially the bounds in SLT hold for any size of the dataset and of the architecture. On the contrary, here we have no control on finite-size corrections, since our results hold only in the limit of large $P$ and $N_{\textrm{out}}$. 
(ii)
As observed in Fig.~\ref{fig:2}, there are cases in which the bound starts to hold only 
after a threshold $P_\mathrm{t} \approx 10^4-10^5$, which the theory does nothing to predict.
However, our numerical experiments (e.g.~those in Fig.~\ref{fig:3}) show
that the threshold is rather small in many empirically relevant cases.
(iii) A Gaussian approximation was
performed at the level of the replicated partition function (for details see appendix \ref{AppA}). Despite similar assumptions have been successfully employed in the past to study the statistical physics of kernels and of random feature models (where excellent agreement with numerics has also been found) \cite{PhysRevLett.82.2975,canatar2021}, we can not guarantee that this approximation is always quantitatively correct.
(iv) These results hold for regression only.
It would be interesting to understand whether the same non-rigorous tools could also be used to study classification problems \cite{loureiro2021learning}. 

Some of these drawbacks may be addressed by a more rigorous approach,
for instance by the use of random matrix theory to avoid replicas
(as done for instance in the case of kernel learning in \cite{pmlr-v119-bordelon20a, 10.2307/26542784, NEURIPS2020_b367e525, goldt2020gaussian, hu2020universality}).
This may shed light on the dependence of the GG on the input dimension $D$,
which we have shown numerically in the lazy training regime to saturate exponentially 
to the mean field bound.

In conclusion, we would like to point out an apparently intriguing consequence of our findings: since our mean field bound suggests a (linear or sub-linear) degradation of the generalisation performance with the last layer size $N_{\textrm{out}}$, we might be led to surmise that, to improve generalisation performance, it may be
convenient to design architectures with a small last layer.
A more systematic investigation on state-of-the-art architectures is needed
to understand whether this insight may lead to design more performant deep neural networks in the future.

\acknowledgements

The authors would like to thank V. Erba, B. Loureiro and M. Pastore for feedback on the manuscript and for pointing out recent related work. R. P. would like to thank her colleagues at Bocconi University for discussions. P. R. acknowledges funding from the Fellini program under the H2020-MSCA-COFUND action, Grant Agreement No. 754496, INFN (IT).

\appendix

\section{Sketch of the replica symmetric calculation}
\label{AppA}
We now discuss the salient aspects of the replica calculation. Further technical details can be found in appendix \ref{AppC}.
Our goal is to evaluate the generalisation and training errors defined in Eqs. \eqref{epsilong}, \eqref{epsilont} for arbitrary teacher function $f_{\textrm{T}}(\mathbf{x})$ and using as a student a function of the form \eqref{student}.

\subsection*{Replicated partition function}

In order to evaluate these observables, one introduces a Gibbs distribution $p_G(\mathbf{v}) = \frac{1}{Z}e^{-\beta \mathcal{L}(\mathbf{v})}$, where $\mathcal{L}$ is the effective Hamiltonian defined in Eq. \eqref{MSEloss} and $\beta$ can be thought as the inverse of an effective temperature. The partition function $Z$ is given by: 

\begin{equation}
    Z = \int  \text{d}^N \,\mathbf{v}  e^{- \frac{\beta}{2}  \sum_{\mu}^P  \left( f_{\textrm{T}}(\mathbf{x}^\mu) - f_{\textrm{S}}(\mathbf{v}, \mathbf{x}^\mu) \right)^2 -\frac{\beta}{2} \lambda \Vert\mathbf{v}\Vert^2} 
\end{equation}
In the $\beta \to \infty$ limit, the Gibbs measure is dominated by the minimum of the Hamiltonian, which corresponds to the minimum of the training loss. 
As for many other problems in disordered systems, meaningful results can be obtained only by {\it quenched} averages, that is by averaging the logarithm of the partition function over all the possible realisations of the training set $\mathcal T$. Physically, this amounts to optimising first the student weights for any given instance of
the dataset $\mathcal T$ and then averaging over all dataset realisations\cite{Gardner_1988}. In order to perform this quenched average we exploit the standard replica method \cite{mezard1987},

\begin{equation}
\langle \log Z \rangle_{\mathcal{T}} = \lim_{m \to 0} \frac{\langle Z^m\rangle_{\mathcal{T}} -1}{m}\,,
\label{replica}
\end{equation}

where one firstly computes the average of $Z^m$ for an integer number of replica $m$ and only later one performs the analytical continuation to $m \rightarrow 0$. 
Since the inputs are drawn as iid variables, the integral over the training set factorises to yield
\begin{equation}
\begin{split}
 \langle Z^m \rangle_{\mathcal{T}} \!&=\!\!
\int \!\prod_{a=1}^m \! \text{d}^N \!\mathbf{v}^a  e^{ - \frac{\beta}{2} \lambda\! \sum_{a}^m \Vert\mathbf{v}^a\Vert^2 }  \\ & \qquad \qquad \times \!\left[ \int\! \text{d}^D\! x\, \rho(\mathbf{x}) e^{- \frac{\beta}{2}\! \sum_a^m (q^a)^2} \right]^P
\end{split}
\label{eq:gauss}
\end{equation}
 where $a$ is a replica index and we introduced a set of auxiliary random variables $q^a \equiv f_{\text{T}}(\mathbf{x}) - \mathbf{v}^a \cdot \pmb{\phi} (\mathbf x)$ with
mean $\mu_q^a (\{\mathbf v^a\})= \langle q^a \rangle_{\rho}$ and covariance matrix:
\begin{equation}
    \begin{split}
        Q_{ab}(\{\mathbf v^a\}) = \langle q^a q^b \rangle_{\rho} =   T \!+\left(\mathbf{v}^a\right)^T \mathbf{\Phi}\; \mathbf{v}^b \!-\!  \mathbf{J}^T\cdot (\mathbf{v}^a\! +\! \mathbf{v}^b)\,,
    \end{split}
\label{eq:cov}
\end{equation}
(with $\mathbf{J}$, $\mathbf{\Phi}$ and $T$ given by Eq. \eqref{JphiT}). To proceed further we note that each of the random variables $q^a$ is the sum of $N$ random variables. For a large last layer size $N$ and input dimension $D$ we approximate their probability distribution with a multivariate Gaussian with mean $\mu_q^a$ and covariance matrix $Q_{ab}$ (an order parameter which measures the overlap between replica $a$ and $b$). 
This allows us to perform the integration in the square brackets in Eq. \eqref{eq:gauss} to get 
\begin{equation}
\begin{split}
&\int\! \text{d}^D\! x\, \rho(\mathbf{x}) e^{- \frac{\beta}{2}\! \sum_a^m (q^a)^2} = \int \prod_a^m dq^a e^{-\frac{\beta}{2}	\sum_a^m (q^a)^2} \times\\
&\;\;\;\;\;\;\;\;\;\times\int d^D x \rho(\mathbf x) \; \delta \left( q_a - \sum_\alpha^{N} v_\alpha^a \phi_\alpha(\mathbf{x})+f_{\textrm T}(\mathbf x) \right)\\
&\simeq \int \prod_a^m d q^a  \,\frac{e^{-\frac{\beta}{2}\sum_a (q^a)^2 -\frac{1}{2}\sum_{a b} (q^a)^T Q^{-1}_{a b} q^b  }}{\sqrt{(2\pi)^{m} \;\det \mathbf{Q} } } =\\
&= \bigg(\det(\mathbb{I} + \beta \mathbf{Q})\bigg)^{-\frac{1}{2}} 
\end{split}
\label{eq:gauss2}
\end{equation}
where we have assumed  $\mu_q^a=0$ without loss of generality \cite{canatar2021}.\\
It is worth noticing that this Gaussian approximation is crucial in order to make progress, but rather uncontrolled, as we lack a formal result demonstrating its validity. 
Nonetheless similar non-rigorous approximations are quite standard in the literature on kernel learning, which include the seminal work on support vector machines by Dietrich, Opper and Sompolinsky \cite{PhysRevLett.82.2975}, more recent findings on kernel regression \cite{canatar2021} or works on the so-called random feature model, where this approximation goes under the name of \emph{Gaussian equivalence principle} \cite{PhysRevX.10.041044}.
These ideas imply that the $N$ feature maps $\phi_\alpha(\bf x)$ are somehow mutually weakly correlated, an assumption deemed resonable 
for a large class of relevant architectures in the thermodinamic limit of large $N$ and $D$ with finite $N/D$ \cite{PhysRevX.10.041044}.

The integration of Eq. (\ref{eq:gauss}) over the weights ${\bf v}^a$, while rather convoluted, now follows a standard replica scheme as detailed in the more technical appendix \ref{AppC}. Thanks to Eq. \eqref{eq:gauss2} and making use of the identities
\begin{equation}
\begin{split}
1 \!&=\! \int \!dQ_{a b} \,\delta\left (Q_{a b} -  \langle q^a q^b \rangle_{\rho} \right) \\ & \!=\!\int dQ_{a b}\frac{d\hat{Q}_{a b}}{2\pi} \,e^{-i \hat{Q}_{a b} \left (Q_{a b} -  \langle q^a q^b \rangle_{\rho} \right)}
\end{split}
\label{ide}
\end{equation} 
one may express Eq. \eqref{eq:gauss} as an integral over the replica order parameter $Q_{ab}$ and its conjugated variables $\hat{Q}_{ab}$. 
Working in the replica symmetric ansatz, which holds for our convex problem, $Q_{ab} = Q_0 \delta_{ab} + Q (1- \delta_{ab} )$ (the same symmetry holding for the conjugated variable), one may compute the leading, linear order contribution in $m$ which determines the limit \eqref{replica}, to get
\begin{equation}
\langle Z^m \rangle_{\mathcal{T}} \approx \int dQ_0 dQ d\hat{Q}_0 d\hat{Q} \,\,e^{-\frac{mP}{2}S_\beta(Q_0, Q, \hat{Q}_0, \hat{Q})}
\end{equation} 
with the rather complicated expression for the action $S_\beta(Q_0, Q, \hat{Q}_0, \hat{Q})$ given by Eq. (\ref{action}) of appendix \ref{AppC}. 

We can finally solve this last integral by saddle-point method in the large $P$ limit. One has to solve a set of four saddle-point equations to find the action minimum that determines the two order parameters of the problem and their conjugated variables (see appendix \ref{AppC}). In the $\beta \to \infty$ limit the order parameter has the rather simple saddle-point solution
\begin{equation}
Q^* = Q_0^* =  \frac{T - P \mathbf{J}^T \left(2\kappa\lambda \mathbb{1} +P\mathbf{\Phi}\right)\bf{G}^{-2}\mathbf{J}}{1-P\text{Tr}[\mathbf{\Phi}^2 \bf{G}^{-2}] }
\label{Q0}
\end{equation}
where the variable $\kappa$ and the invertible matrix $\bf{G}$ have been introduced in the main text.

\subsection*{Generalisation and training errors}
The generalisation $\epsilon_g$ and training $\epsilon_t$ errors can be easily related to the saddle-point replica order parameters. To see this, first consider the generalisation error. From Eqs. \eqref{epsilong} and \eqref{student} in the main text
\begin{equation}
\begin{split}
\label{eq34}
        \epsilon_{\text{g}} &= \int
        d^Dx \rho(\mathbf{x}) \left(f_T(\mathbf{x}) - \mathbf{v}^*\cdot\pmb{\phi}(\mathbf{x}) \right)^2\\
        &= T -2 \mathbf{J}^T\mathbf{v}^* + {\mathbf{v}^*}^T \mathbf{\Phi} \mathbf{v}^*=Q^*\,.
\end{split}
\end{equation}
where in the last equality we used Eq. \eqref{eq:cov} and the replica symmetric ansatz.
In the limit $P\rightarrow \infty$, it is easy to show that $\kappa \rightarrow 1$ and $\mathbf{G} \sim P \mathbf{\Phi}$ so that the generalisation error converges to the \emph{residual generalisation error} introduced in the main text, Eq. \eqref{epsR} 
\begin{equation}
\epsilon_{\text{g}} \rightarrow \epsilon_{\text{g}}^{\textrm{R}} = T - \mathbf{J}^T \mathbf{\Phi}^{-1}\mathbf{J}\,.     
\end{equation}
Notice that this result is not surprising and it provides a first consistency check of our replica mean field theory: one could also obtain it by directly minimising Eq. \eqref{eq34} with respect to the parameters $\mathbf{v}$. In fact, $\partial_{\mathbf{v}} \epsilon_{\text g} = - 2 \mathbf{J}^T + 2 \mathbf{\Phi} \mathbf{v}$ implies 
$\mathbf{v}^* = \mathbf{J}^T\mathbf{\Phi}^{-1}$, so that $\epsilon_{\text g}(\mathbf{v}^*) \equiv \epsilon_{\text g}^{\text R} = T - \mathbf{J}^T \mathbf{\Phi}^{-1}\bf{J}$.
Furthermore, by isolating the residual generalisation error in Eq. \eqref{Q0}, we finally find the compact formula for the generalisation error quoted in Eq. \eqref{traintesteq}  of the main text.

Using the theory developed so far, we can also have access to the average value of the training error. We notice that the average training error is by definition the average loss function defined in Eq. \eqref{MSEloss}, evaluated in $\lambda = 0$ (up to a factor $P$). This means that one can extract its value by evaluating the action on the saddle point solution and performing the limit $\beta \rightarrow \infty$, i.e.: 
\begin{equation}
    \begin{split}
        \epsilon_{\text t} = \lim_{\beta \to \infty} \frac{1}{\beta}\left. S_\beta (Q^*,Q_0^*,\hat{Q}^*,\hat{Q}_0^*)\right|_{\lambda=0} \,.
    \end{split}
\end{equation}
After some straightforward but lengthy algebraic manipulations, one recovers the training error given in Eq. \eqref{traintesteq}  of the main text.

\section{Numerical experiments details}
\label{AppB}

In this section we give a detailed report of all our numerical procedures. The code to replicate our experiments can be found at \textit{https://github.com/rosalba-p/Generalisation$\_$DNN}. 

\subsection*{Teacher-student architectures}

\paragraph*{Student architectures.} We considered six types of student architectures: two toy networks with one- and two-hidden layers (size of the second hidden layer $N_{\rm hid} = 200$) and three state-of-the-art convolutional ones (ResNet18, DenseNet121 and VGG11). The toy architectures have fully connected layers and ReLu activation functions at every layer but the last; the convolutional networks are the standard PyTorch \cite{PyTorch} models modified to yield a scalar output suitable for regression through a last fully connected linear layer with parameters $\mathbf{v}$, instead of the $LogSoftMax$ that is employed for classification. All these architectures have several convolutional layers before a last fully connected one that counts respectively $N_{\text{out}} =512,1024,4096$ hidden units. The total number of trainable parameters in the first two networks (weights and biases) is approximately 10 million, while vgg11 counts 10 times as many.
\paragraph*{Teacher architectures and inputs for toy DNNs.} Each linear or quadratic teachers (see Results) is defined by a random uniform vector $\mathbf{t} \in \mathbb{R}^D$ of unitary norm. 1HL teachers are defined by parameters $q_\alpha \in  \mathbb{R} $and $\mathbf{S}_{\alpha} \in \mathbb{R}^D$  ($\alpha=1,\ldots,M=200$). They are drawn from a normal distribution with zero mean and variance (respectively) $1/M$ and $1/D$. 
Inputs $\mathbf{x} \in \mathbb{R}^D$ are also drawn from normal distribution with zero mean and unit variance.

\subsection*{Learning and generalization}
\begin{figure}[t!]
    \centering
    \includegraphics[width = 0.42\textwidth]{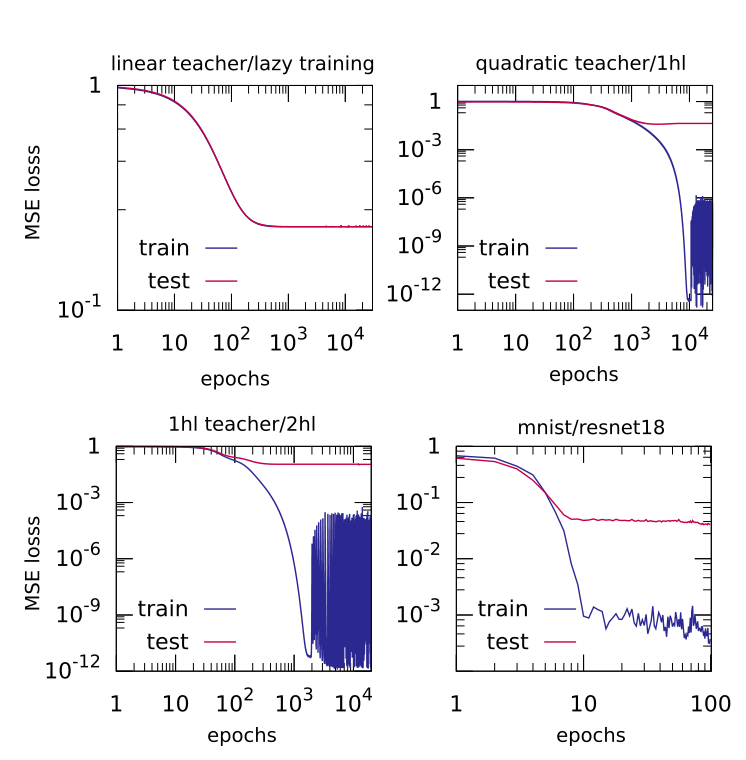}
    \caption{Train (blue) and test (magenta) loss of different teacher/student tasks as a function of the training epochs. The test loss reaches a plateau even if the training loss is noisy, due to the different order of magnitude reached by the two. It is worth remarking that the losses are normalised with the trivial predictor and therefore at epoch 1 are $O(1)$.}
    \label{fig:loss}
\end{figure}
\paragraph*{Learning algorithm.} All the architectures are trained with Adam optimiser \cite{Adam_2015} and a small weight decay ($w_d = 10^{-5}$), while the learning rate $\alpha$ is set to $10^{-3}$ for the toy architectures and $10^{-4}$ for the convolutional ones. The weight decay $w_d$ is related to the regularisation parameter $\lambda$ via $w_d = \lambda \alpha$, and we have verified that our results do not change quantitatively by varying $\lambda$ in a reasonable range ($\lambda \leq 10^{-1}$).
The regression loss employed is the standard mean squared error,
\begin{equation}
    \text{MSE} = \frac{1}{P} \sum_{\mu=1}^{P} \left( f_{\text{T}}(x^\mu) - f_{\text{S}}(x^\mu) \right)^2
\end{equation} 
For the MNIST dataset, we consider the labels as integers: $f_{\text{S}}(x^\mu) = [0,1,2,3,4,5,6,7,8,9]$, and the MSE loss is computed as in the other cases. 
\paragraph*{Estimation of the generalisation gap.} We first bring the training procedure to convergence, i.e. we ensure that the train and test loss have reached a plateau. 
For synthetic datasets, this requires a large number of training epochs ($3 \cdot 10^4$), while MNIST is learnt faster (100 epochs). In one epoch the network is fed all the dataset, arranged in minibatches only when full batch learning is prohibited by memory limitations. Once the train loss is steady with respect to the test loss, we retrieve the generalisation gap as the average over the last 100 epochs (50 for MNIST). The size of the train set is $P_{\text{test}}=10^4$ is all cases. \\
Several plots of train and test loss vs the number of epochs are shown in fig \ref{fig:loss}: for different teacher-student pairs the plateau in the test loss is always reached: even if some noise is still visible in the train loss, its oscillations are too small to affect the test loss and therefore the generalisation gap.   
\subsection*{Trivial predictor.} To compare results obtained from different teacher/student pairs, we need to normalise the loss by its natural scale, i.e. the trivial predictor $T$ defined in Eq. \eqref{JphiT}. By doing this, we make sure that the train and test loss are always of $O(1)$ for a random architecture (i.e. at epoch 0). $T$ is a property of the dataset, and its computation changes accordingly.

\begin{figure}[t]
    \centering
    \includegraphics[width = 0.42\textwidth]{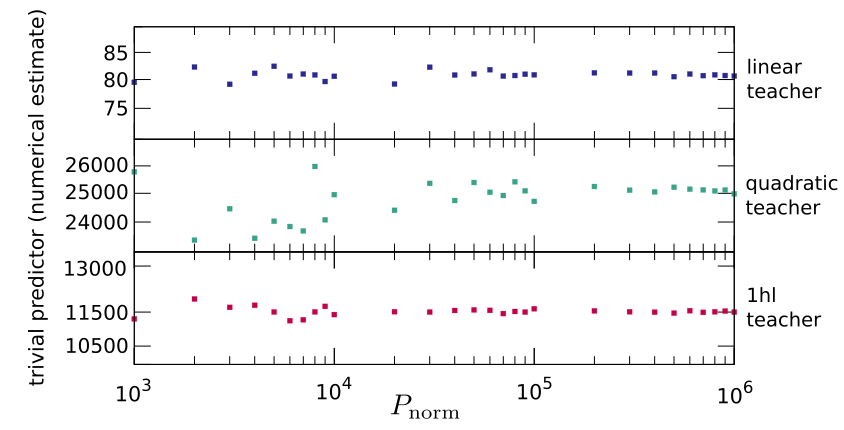}
    \caption{Numerical estimate of the trivial predictor $T$ as a function of the number of points $P_{\text{norm}}$ used in the approximation. This values describe a single realisation of a random teacher function (linear, quadratic, 1hl). $T$ converges to a fixed value around $P_{\text{norm}} = 10^{6}$, that is the value chosen for all our numerical experiments.}
    \label{fig:trivialpred}
\end{figure}

\paragraph*{MNIST.}
The MNIST dataset is intrinsically suitable for classification problems, since it has 10 classes with discrete labels, that are simply the first 10 integers. For this reason, the integral that describes the trivial predictor here becomes a summation over the classes. Another simplification can be made considering that MNIST is balanced over the training labels: the $P = 6 \cdot 10^{4}$ training examples are equally distributed over the 10 classes. In this case the computation of the trivial predictor is simple and can be performed analytically:
\begin{equation}
    T = \frac{1}{P} \sum_{n = 0}^{9} \frac{P}{\text{\# classes}} n^2 =  \frac{1}{\text{\# classes}} \sum_{n = 0}^{9} n^2 = 28.5
\end{equation}

\paragraph*{Synthetic datasets.}
For the other teachers that we employed $T$ cannot be computed analytically. We therefore perform a numerical estimation in the following approximation: 
\begin{equation}
    T \sim \frac{1}{P_{\text{norm}}}\sum_{\mu=1}^{P_{\text{norm}}} f^2_{\text{T}}(x^\mu)
\end{equation}
Operatively, we draw an independent data sample of $P_{\text{norm}}$ elements $\lbrace x^\mu \rbrace_{\mu = 1 \ldots P_{\text{norm}}}$, and average the respective squared true labels $f^2_{\text{T}}(x^\mu)$.
$P_{\text{norm}}=10^{6}$ was chosen as a safe compromise between computational time and consistency of the estimate. The convergence of $T$ to a fixed value when $P_{\text{norm}}$ grows is shown in Figure \ref{fig:trivialpred}.

\subsection*{Higher dimensional synthetic inputs.}
\begin{figure}[t!]
    \centering
    \includegraphics[width = 0.49\textwidth]{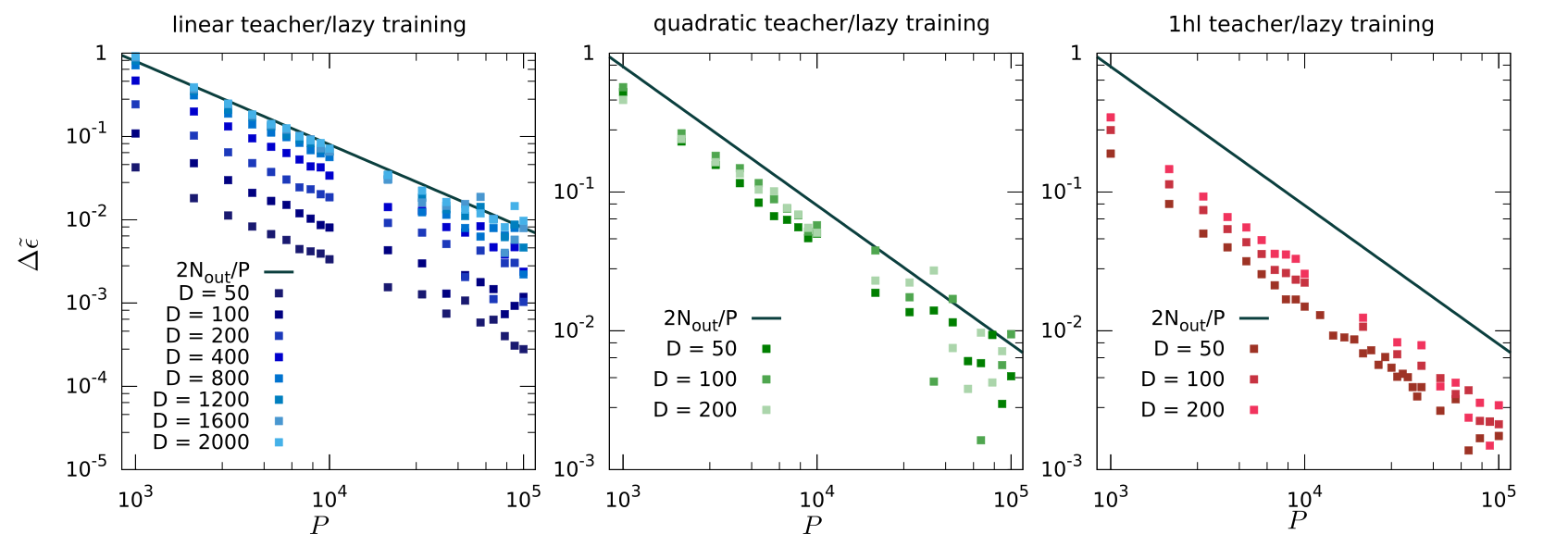}
    \caption{Generalisation gap of a lazy training architecture as a function of the number of examples in the training set for different input sizes $D$. From left to right the teacher functions are respectively linear, quadratic and 1hl. The size of the hidden layer is kept fixed to $N_{\text{out}}= 400$.}
    \label{fig:SM_Dscaling}
\end{figure}
In the main text we show the generalisation gap scaling for fixed input size $D = 50$, with the exception of the inset in Fig 1. Here we report the full analysis of higher dimensional input size and clarify the procedure used to obtain the inset.
In Figure \ref{fig:SM_Dscaling} we plot the generalisation gap of of a lazy training architecture learning the three function considered in the main text (linear, quadratic and 1hl), as a function of the number of training examples and for different sizes of the input $D$. Increasing $D$ does not change the asympthotic behaviour of the curves, but has the effect of pushing them closer to the bound given the same trainset size $P$. To assert how fast the curves approach the bound, we have performed a linear fit of the generalisation gap as a function of $P$: 
\begin{equation}
    \log \Delta \tilde{\epsilon} = m \cdot  \log P + y_0 
\end{equation}
Note that for the bound $m \equiv -1$ and $y_0 \equiv \log 2N_{\text{out}}$. For different values of $D$, it is verified that $m \sim -1$, while $y_0 - 2N_{\text{out}}$ approaches 0 exponentially fast (see also inset in Fig. 1 of the main text).
%

\section{Replica symmetric ansatz and saddle point equations}
\label{AppC}

Thanks to the Gaussian approximation (\ref{eq:gauss2}) and making use of the identities (\ref{ide}), the replicated partition function  (\ref{eq:gauss}) may be recasted in the following form
\begin{equation}
\begin{split}
&\langle Z^m \rangle_{\mathcal{T}} = \int dQ_{a b}d\hat{Q}_{a b}\; K(\beta, Q)^{-\frac{P}{2}} e^{iQ_{a b}\hat{Q}_{a b}} \times \\
&\times\int d^N\mathbf{v}^a e^{-\frac{\beta \lambda}{2}\sum_{a} \Vert\mathbf{v}^a \Vert^2-i\sum_{a\leq b} \hat{Q}_{a b} \left(T +\left(\mathbf{v}^a\right)^T \mathbf{\Phi}\; \mathbf{v}^b - \mathbf{J}^T(\mathbf{v}^a+\mathbf{v}^b) \right)}
\end{split}
\label{finalZ}
\end{equation}
where
\begin{equation}
K(\beta, {\bf Q}) \equiv \det(\mathbb{I} + \beta \mathbf{Q})\,.
\end{equation}
The idea is now to explicitly perform the integration over the replicated weights $\mathbf v^a$ and to evaluate the integrals over $Q_{a b}$ and $\hat{Q}_{a b}$ with the saddle-point method. As it occurs in standard spin glass models, we are left with the complication of performing the tricky limit $m\rightarrow 0$, which may lead to the breaking of the so-called replica symmetry of the matrix $Q_{a b}$. However in this specific case the underlying optimisation problem that we are dealing with is convex and this guarantees that the simplest replica symmetric ansatz for the matrix $Q_{a b}$ will provide the exact solution to the saddle-point equations in the limit $m\rightarrow 0$ \cite{PhysRevLett.82.2975}.  

The assumption of replica symmetry amounts to require that the matrices $Q_{a b}$/$\hat{Q}_{a b}$ take the following form:
\begin{equation}
Q_{a b} = 
\begin{cases}
      Q_0 \quad a = b \\
      Q\,\,   \quad a \neq b
\end{cases}
\quad
\hat{Q}_{a b} = 
\begin{cases}
      \hat{Q}_0 \quad a = b \\
      \hat{Q}\,\,   \quad a \neq b\,.
\end{cases}
\end{equation}
With a slight abuse of notation, we can specify our analysis to the replica symmetric ansatz already at the level of Eq. \eqref{finalZ}, thus obtaining:
\vspace{0.1cm}
\begin{widetext}
\begin{equation}
    \begin{split}
        \langle Z^m \rangle_{\mathcal{T}} =& \int dQ_0\;dQ\;d\hat{Q}_0\;d\hat{Q}\; K^{-\frac{P}{2}}e^{im\left(\hat{Q}_0(Q_0-T)+\frac{m-1}{2}\hat{Q}(Q-T)\right)} \times \\
        &\times \int d^N \mathbf{v}^a e^{-\frac{\beta \lambda}{2}\sum_{\alpha,a} (v_\alpha^a)^2- i\left(\hat{Q}_0-\frac{\hat{Q}}{2}\right)\sum_a(\mathbf{v}^a)^T \mathbf{\Phi} \mathbf{v}^a}e^{ -i\frac{\hat{Q}}{2}\sum_{a b}(\mathbf{v}^a)^T\mathbf{\Phi}\mathbf{v}^b + 2i\left(\hat{Q}_0 -(1-m)\frac{\hat{Q}}{2}\right)\sum_a \mathbf{J}^T\mathbf{v}^a} \,.
    \end{split}
\end{equation}
By performing the Gaussian integrals over the replicated weights and using standard mathematical manipulations, we finally get the following result:
\begin{equation}
    \begin{split}
        &\langle Z^m \rangle_{\mathcal{T}} = \int dQ_0\;dQ\;d\hat{Q}_0\;d\hat{Q}\; K^{-\frac{P}{2}}e^{-\frac{m}{2}\left(\hat{Q}_0(Q_0-T)+(m-1)\hat{Q}(Q-T)\right)}\times\\ 
&\;\;\;\;\;\;\;\times (2\pi)^{\frac{m}{2}}e^{-\frac{m-1}{2}\text{Tr}\left[\log\left(\beta\lambda\mathbb{1}-(\hat{Q}_0-\hat{Q})\mathbf{\Phi} \right)\right]-\frac{1}{2}\text{Tr}\left[\log\left(\beta\lambda\mathbb{1}-(\hat{Q}_0-(1-m)\hat{Q})\mathbf{\Phi}\right)\right]} e^{\frac{m}{2}\left(\hat{Q}_0-(1-m)\hat{Q}\right)^2 \mathbf{J}^T\left(\beta\lambda\mathbb{1}-(\hat{Q}_0-(1-m)\hat{Q})\mathbf{\Phi}\right)^{-1}\mathbf{J}}   
    \end{split}
\end{equation}
\end{widetext}

At this point we have managed to integrate both on the dataset $\mathcal T$ and on the replicated weights. To find the solution of the last integrals over the order parameters, we exploit the saddle-point method, which will deliver the exact solution in the limit $P\rightarrow \infty$.

\subsection*{Replica symmetric ansatz}
First we notice that assuming replica symmetry, the contribution $K$ to the partition function simplifies in the following way: 
\begin{equation}
    \begin{split}
        K(\beta,Q,Q_0) &= \det[\mathbb{I}+\beta\mathbf{Q}]\\
        &\!=\! (1 \!+\!\beta(Q_0-Q))^{m-1}(1 \!+\!\beta(Q_0 \!-\! (1 \!-\! m)Q))\,.
    \end{split}
\end{equation}

In order to extract the $m\rightarrow 0$ limit, we have to keep only those terms that are linear in $m$. This is easily done term by term:
\begin{equation}
    \begin{split}
        &e^{-\frac{P}{2}\log\left((1+\beta(Q_0-Q))^{m-1}(1+\beta(Q_0-(1-m)Q))\right)}\\&\approx e^{-\frac{P}{2} \; m \left( \frac{\beta Q}{1+\beta(Q_0-Q)}+\log(1+\beta(Q_0-Q))  \right)}\,,
    \end{split}
\end{equation}
\vspace{0.1cm}
\begin{equation}
    \begin{split}
        e^{-\frac{m}{2}\left(\hat{Q}_0(Q_0-T)+(m-1)\hat{Q}(Q-T)\right)}\approx e^{-\frac{m}{2}\left(\hat{Q}_0(Q_0-T)-\hat{Q}(Q-T)\right)}\,,
    \end{split}
\end{equation}
\begin{equation}
    \begin{split}
       & e^{-\frac{m-1}{2}\text{Tr}\left[\log\left(\beta\lambda\mathbb{1}-(\hat{Q}_0-\hat{Q})\mathbf{\Phi}\right)\right]-\frac{1}{2}\text{Tr}\left[\log\left(\beta\lambda\mathbb{1}-(\hat{Q}_0-(1-m)\hat{Q})\mathbf{\Phi}\right)\right]} \\&\approx e^{-\frac{m}{2}\text{Tr}\left[\log\left(\beta\lambda\mathbb{1}-(\hat{Q}_0-\hat{Q})\mathbf{\Phi}\right)\right]+\frac{m\hat{Q}}{2}\text{Tr}\left[\mathbf{\Phi}\left(\beta\lambda\mathbb{1}-(\hat{Q}_0-\hat{Q})\mathbf{\Phi}\right)^{-1}\right] }\,,
    \end{split}
\end{equation}
and
\begin{equation}
    \begin{split}
&e^{\frac{m}{2}\left(\hat{Q}_0-(1-m)\hat{Q}\right)^2 \mathbf{J}^T\left(\beta\lambda\mathbb{1}-(\hat{Q}_0-(1-m)\hat{Q})\mathbf{\Phi}\right)^{-1}\mathbf{J}}  \\&\approx e^{\frac{m}{2}\left(\hat{Q}_0-\hat{Q}\right)^2 \mathbf{J}^T\left(\beta\lambda\mathbb{1}-(\hat{Q}_0-\hat{Q})\mathbf{\Phi}\right)^{-1}\mathbf{J} }  \,.
    \end{split}
\end{equation}
As such, the leading contribution to the average replicated partition function in the $m \rightarrow 0$ limit reads:
\begin{widetext}
\begin{equation}
    \begin{split}
        \langle Z^m \rangle_{\mathcal{T}} \sim&\int dQ_0\;dQ\;d\hat{Q}_0\;d\hat{Q}\;
        e^{-\frac{P}{2} \; m \left( \frac{\beta Q}{1+\beta(Q_0-Q)}+\log(1+\beta(Q_0-Q))  \right)} e^{-\frac{m}{2}\left(\hat{Q}_0(Q_0-T)-\hat{Q}(Q-T)\right)}\\&\qquad \times e^{-\frac{m}{2}\text{Tr}\left[\log\left(\beta\lambda\mathbb{1}-(\hat{Q}_0-\hat{Q})\mathbf{\Phi}\right)\right]+\frac{m\hat{Q}}{2}\text{Tr}\left[\mathbf{\Phi}\left(\beta\lambda\mathbb{1}-(\hat{Q}_0-\hat{Q})\mathbf{\Phi}\right)^{-1}\right]} e^{\frac{m}{2}\left(\hat{Q}_0-\hat{Q}\right)^2 \mathbf{J}^T\left(\beta\lambda\mathbb{1}-(\hat{Q}_0-\hat{Q})\mathbf{\Phi}\right)^{-1}\mathbf{J}}        
    \end{split}
\end{equation}
\end{widetext}
and by rescaling the parameters $\hat{Q}\to P\hat{Q}$ and $\hat{Q}_0\to P\hat{Q}_0$, we recast the partition function in the form $\langle Z^m \rangle_{\mathcal{T}}=\int e^{-\frac{mP}{2}S_\beta}$, where the action $S_\beta$ is defined as:
\begin{equation}
    \begin{split}
       S_\beta= &
        \;\frac{\beta Q}{1+\beta(Q_0\!-\!Q)}+\log(1+\beta(Q_0\!-\!Q))
        + \hat{Q}_0(Q_0-T) +\\
&-\hat{Q}(Q-T)+\frac{1}{P}\text{Tr}\left[\log\left(\beta\lambda\mathbb{1}-P(\hat{Q}_0-\hat{Q})\mathbf{\Phi}\right)\right]+\\
&-\hat{Q}\text{Tr}\left[\mathbf{\Phi}\left(\beta\lambda\mathbb{1}-P(\hat{Q}_0-\hat{Q})\mathbf{\Phi}\right)^{-1}\right]+\\
        &-P\left(\hat{Q}_0-\hat{Q}\right)^2 \mathbf{J}^T\left(\beta\lambda\mathbb{1}-P(\hat{Q}_0-\hat{Q})\mathbf{\Phi}\right)^{-1}\mathbf{J}       
    \end{split}
\label{action}
\end{equation}
\subsection*{Saddle point equations}
We can now move to the derivation of the saddle-point equations.  Firstly we notice that direct differentiation with respect to $Q$ and $Q_0$ allows to find the explicit expressions for $\hat Q$ and $\hat{Q}_0$:
\begin{equation}
        0=\frac{\partial S_\beta}{\partial Q} = - \hat{Q} + \frac{\beta^2 Q}{(1+\beta(Q_0-Q))^2}  
\end{equation}
which implies
\begin{equation}
\hat{Q} = \frac{\beta^2 Q}{(1+\beta(Q_0-Q))^2} 
\label{SP1}
\end{equation}
and
\begin{equation}
       0= \frac{\partial S_\beta}{\partial Q_0} = \hat{Q}_0 + \frac{-\beta Q (\beta)}{\left(1+\beta(Q_0-Q)\right)^2} + \frac{\beta}{1+\beta(Q_0-Q)} 
\end{equation}
which gives
\begin{equation}
\begin{split}
       & \qquad \hat{Q}_0 = \frac{\beta^2 Q}{(1+\beta(Q_0-Q))^2} - \frac{\beta}{1+\beta(Q_0-Q)}=\\
       &\qquad \qquad = \hat{Q} - \frac{\beta}{1+\beta(Q_0-Q)}
    \end{split}
    \label{SP2}
\end{equation}
Let us look at the derivative of the action w.r.t. $\hat{Q}$:
\begin{equation}
    \begin{split}
        0=&\frac{\partial S_\beta}{\partial \hat{Q}} = -(Q-T) + \frac{1}{P}\partial_{\hat{Q}}\text{Tr}\left[\log\left(\tilde{\mathbf{G}}\right)\right] -\text{Tr}\left[\mathbf{\Phi} \tilde{\mathbf{G}}^{-1}\right] + \\
&-\hat{Q} \partial_{\hat{Q}} \text{Tr}\left[\mathbf{\Phi} \tilde{\mathbf{G}}^{-1}\right] 
\!+\! P \mathbf{J}^T \frac{P\mathbf{\Phi}(\hat{Q}_0 \!-\!\hat{Q})^2 \! \!+\! 2\tilde{\mathbf{G}}(\hat{Q}_0 \!-\!\hat{Q})}{\tilde{\mathbf{G}}^2}\mathbf{J}
\end{split}
\end{equation}
that gives
\begin{equation}
    \begin{split}
       &Q = T + \frac{1}{P}\partial_{\hat{Q}}\text{Tr}\left[\log\left(\tilde{\mathbf{G}}\right)\right] -\text{Tr}\left[\mathbf{\Phi} \tilde{\mathbf{G}}^{-1}\right] +\\
&\;\;\;- \hat{Q} \partial_{\hat{Q}} \text{Tr}\left[\mathbf{\Phi} \tilde{\mathbf{G}}^{-1}\right] 
        + P \mathbf{J}^T \frac{P\mathbf{\Phi}(\hat{Q}_0-\hat{Q})^2+2\tilde{\mathbf{G}}(\hat{Q}_0-\hat{Q})}{\tilde{\mathbf{G}}^2}\mathbf{J}\\ & =
        T  \!+\! P\,\hat{Q}\; \text{Tr}[\mathbf{(\Phi} \tilde{\mathbf{G}}^{-1})^2] 
         \!+\! P \mathbf{J}^T \frac{P\mathbf{\Phi}(\hat{Q}_0 \!-\!\hat{Q})^2 \! \!+\! 2\tilde{\mathbf{G}}(\hat{Q}_0 \!-\!\hat{Q})}{\tilde{\mathbf{G}}^2}\mathbf{J}
    \end{split}
    \label{SP3}
\end{equation}
where for convenience we have defined the $N\times N$ matrix $\Tilde{\mathbf{G}} = \beta\lambda\mathbb{1}-P(\hat{Q}_0-\hat{Q})\mathbf{\Phi}$. Finally we obtain the saddle point equation for  $\hat{Q}_0$:
\begin{equation}
    \begin{split}\!
        0=&\frac{\partial S_\beta}{\partial \hat{Q}_0} = (Q_0-T) + \frac{1}{P}\partial_{\hat{Q}_0}\text{Tr}\left[\log\left(\tilde{\mathbf{G}}\right)\right] +\\
& - \hat{Q} \partial_{\hat{Q}_0} \text{Tr}\left[\mathbf{\Phi} \tilde{\mathbf{G}}^{-1}\right] 
 \!-\! P \mathbf{J}^T \frac{P\mathbf{\Phi}(\hat{Q}_0 \!-\!\hat{Q})^2 \! \!+\! 2\tilde{\mathbf{G}}(\hat{Q}_0 \!-\!\hat{Q})}{\tilde{\mathbf{G}}^2}\mathbf{J}
\end{split}
\end{equation}
giving us
\begin{equation}
    \begin{split}
    &  Q_0 =  T + \text{Tr}\left[\mathbf{\Phi} \tilde{\mathbf{G}}^{-1}\right]  + P\,\hat{Q}\; \text{Tr}[\mathbf{\Phi} \tilde{\mathbf{G}}^{-1}\mathbf{\Phi} \tilde{\mathbf{G}}^{-1}] 
        \\ &\qquad \qquad + P \mathbf{J}^T \frac{P\mathbf{\Phi}(\hat{Q}_0-\hat{Q})^2+2\tilde{\mathbf{G}}(\hat{Q}_0-\hat{Q})}{\tilde{\mathbf{G}}^2}\mathbf{J}
        \\&\quad\qquad = Q +\text{Tr}\left[\mathbf{\Phi} \tilde{\mathbf{G}}^{-1}\right]
    \end{split}
    \label{SP4}
\end{equation}
Let us now consider the special combination $\kappa = 1 + \beta (Q_0-Q)$. By using Eq. \eqref{SP4}, we obtain:
\begin{equation}
    \begin{split}
        \kappa  = &1 + \beta (Q_0-Q) = 1 + \beta \left( Q +\text{Tr}\left[\mathbf{\Phi} \tilde{\mathbf{G}}^{-1}\right]-Q\right) \\&= 1 +  \beta\;\text{Tr}\left[\mathbf{\Phi} \tilde{\mathbf{G}}^{-1}\right]\,,
    \end{split}
\end{equation}
whereas by considering the difference between Eq. \eqref{SP1} and Eq. \eqref{SP2}, we easily show that the difference $\hat{Q}_0 - \hat{Q}$ depends on $\kappa$ only as:
\begin{equation}
    \begin{split}
        \hat{Q}_0\!-\!\hat{Q} = \hat{Q}\! -\! \frac{\beta}{1\!+\!\beta(Q_0\!-\!Q)} \!-\! \hat{Q} = -\frac{\beta}{1\!+\!\beta(Q_0\!-\!Q)} = -\frac{\beta}{\kappa}
    \end{split}
\end{equation}
By inserting this result into the definition of $\tilde{\mathbf{G}}$ we can define the rescaled matrix $\mathbf{G} = \kappa \tilde{\mathbf{G}}/\beta$:
\begin{equation}
    \begin{split}
        \mathbf{G} =  \frac{\kappa}{\beta}\left[\beta\lambda\mathbb{1}-P(\hat{Q}_0-\hat{Q})\mathbf{\Phi} \right]= \kappa\lambda\mathbb{1}+P\mathbf{\Phi}\,.
    \end{split}
\end{equation}
These observations allow us to show that the new variable $\kappa$ satisfies the following self-consistency equation:
\begin{equation}
    \begin{split}
        \kappa =& 1 + \beta\;\text{Tr} \left[\mathbf{\Phi} \left(\frac{\beta}{\kappa} \mathbf{G}\right)^{-1}\right]=1 + \kappa \text{Tr} \left[\mathbf{\Phi} \mathbf{G}^{-1}\right]
\\=& 1 + \kappa \;\text{Tr} \left[\frac{\mathbf{\Phi}}{\kappa\lambda\mathbb{1}+P\mathbf{\Phi}}\right]
    \end{split}
    \label{kappa}
\end{equation}
and allow to recast the solution of the saddle-point equations in the following very convenient form (notice that from now on the solutions of the saddle-point equations will be indicated with an asterisk):
\begin{equation}
    \begin{split}
        &\hat{Q}^*_0 = \hat{Q}^* - \frac{\beta}{\kappa}\\
        &\hat{Q}^* = \frac{\beta^2Q^*}{\kappa^2}\\
        & Q_0^* = Q^* +\frac{\kappa -1}{\beta}\\ 
        & Q^* =  \frac{T - P \mathbf{J}^T\frac{2\kappa\lambda+P\mathbf{\Phi}}{\mathbf{G}^2}\mathbf{J}}{1-P\text{Tr}[\mathbf{\Phi}^2 \mathbf{G}^{-2}] }\,.
        \end{split}
\end{equation}
It is worth noticing that since $\mathbf{G}$ and $\kappa$ are independent on the inverse temperature $\beta$, the solution for the order parameter $Q^*$ is also independent on the temperature. Moreover, we easily get that $Q^* = Q^*_0$ in the limit $\beta~\rightarrow~\infty$.
\vspace{1.5cm}

\bibliographystyle{ieeetr}
\bibliography{biblio}

\end{document}